\documentclass[runningheads]{llncs}

\usepackage{graphicx}

\begin{document}

\title{Data Privacy with Homomorphic Encryption in Neural Networks Training and Inference\thanks{This work was partially supported by the Norte Portugal Regional Operational Programme (NORTE 2020), under the PORTUGAL 2020 Partnership Agreement, through the European Regional Development Fund (ERDF), within project ``Cybers SeC IP'' (NORTE-01-0145-FEDER-000044). This work has also received funding from the project UIDB/00760/2020.}}
\titlerunning{Data Privacy with HE in NN Training and Inference}

\author{
Ivone Amorim\inst{1}\orcidID{0000-0001-6102-6165}\and
Eva Maia\inst{2}\orcidID{0000-0002-8075-531X} \and
Pedro Barbosa\inst{2}\orcidID{0000-0002-7381-3462}\and 
Isabel Praça\inst{2}\orcidID{0000-0002-2519-9859}
\authorrunning{I. Amorim et al.}}
%
\institute{ PORTIC – Porto Research, Technology \& Innovation Center,
Polytechnic of Porto~(IPP), 4200-374 Porto
Portugal\\ \email{ivone.amorim@portic.ipp.pt}
\and
Research Group on Intelligent Engineering and Computing for Advanced Innovation and Development (GECAD), Porto School of Engineering, Polytechnic of Porto~(ISEP-IPP), 4200-072 Porto, Portugal \\
\email{\{egm,pmbba,icp\}@isep.ipp.pt}}
\maketitle              
\begin{abstract}
The use of Neural Networks (NNs) for sensitive data processing is becoming increasingly popular, raising concerns about data privacy and security. Homomorphic Encryption (HE) has the potential to be used as a solution to preserve data privacy in NN. This study provides a comprehensive analysis on the use of HE for NN training and classification, focusing on the techniques and strategies used to enhance data privacy and security.  The current state-of-the-art in HE for NNs is analysed, and the challenges and limitations that need to be addressed to make it a reliable and efficient approach for privacy preservation are identified. Also, the different categories of HE schemes and their suitability for NNs are discussed, as well as the techniques used to optimize the accuracy and efficiency of encrypted models. 
The review reveals that HE has the potential to provide strong data privacy guarantees for NNs, but several challenges need to be addressed, such as limited support for advanced NN operations, scalability issues, and performance trade-offs. 

\end{abstract}


\section{Introduction}

The emergence of digital technology and the ever-increasing significance of data in the economy have made it crucial to protect personal and sensitive information. As the amount of data being collected and stored grows, so does the risk of privacy breaches and data misuse.  Furthermore, the use of intelligent techniques like Neural Networks (NNs) to extract insights from this data presents another potential vulnerability. Considering this, many privacy-preserving techniques have been proposed, including differential privacy, secure multiparty computation, and homomorphic encryption (HE). Among these, HE has attracted significant attention due to its ability to enable computations on encrypted data without requiring decryption, which can enhance data privacy while still allowing for data analysis. In recent years, HE has been increasingly applied to NNs. By training NNs on encrypted data using HE, sensitive information can be kept private, as the NN can operate on the encrypted data without needing to decrypt it. This can be particularly useful in scenarios where data privacy is a major concern, such as healthcare or finance. However, there are still challenges and limitations to the use of HE in NNs for data privacy, including computational overhead and potential security vulnerabilities.




 Some scientific works have already proposed the use of HE to preserve data privacy in NNs. One of the most relevant works in this area was introduced by Dowlin et al.~\cite{cryptoeprint:2018/462}, who modified a trained NN to operate on encrypted data, creating what they called a CryptoNet. However, this work assumes that the model is trained based on plain data and then this model is adapted for classifying encrypted instances. Other works have emerged to address the limitations of this method and propose new ones~\cite{AIHE_Pulido-Gaytan21}, but few have explored the use of HE for NN training and classification. Furthermore, the current literature in this area is almost exclusively addressed by practitioners looking for suitable implementations. To the best of our knowledge, there are only two publications that provide an overview of the existing approaches to preserve data privacy in NNs using HE: Poulido-Gaytan et al.~\cite{AIHE_Pulido-Gaytan21} and Podschwadt et al.~\cite{AIHE_Podschwadt22}. The former focuses mainly on works that use HE solely in the classification phase, which is the most prevalent approach in the literature. In fact, only one reference in their work addresses the problem of using HE in NN training and inference. On the other hand, Podschwadt's work is centred on deep learning architectures for privacy-preserving machine learning with Fully HE, without including other types of HE schemes like Partial or Somewhat HE.  
Therefore, in this work, we aim to provide a comprehensive analysis of the use of HE for NN training and classification. We will review and analyse the current state-of-the-art in this field and identify the limitations of using homomorphically encrypted data in NNs. We will also analyse existing approaches aimed at addressing these limitations and discuss future research directions in this topic.


\section{Background}\label{sec:back}


Neural Networks (NNs) are weighted directed graphs that are inspired by the way the human brain works \cite{B_Richard88}. 
The nodes of these graphs are usually called neurons. These neurons are grouped in layers and are visualized as being stacked. 
Each of the neurons computes a function over the values of the layer beneath it. 
This function is usually named Activation Function (AF) and is applied to the output of each neuron in a NN, to determine whether the neuron should be activated (output a signal) or not, based on the input it receives. Common activation functions include the sigmoid function, ReLU (Rectified Linear Unit), and tanh (hyperbolic tangent) function. The 
loss function 
is then used to measure NN performance, by calculating the difference between the predicted output and the actual output. The goal of training a NN is to minimize the loss function, typically using an optimization algorithm such as stochastic gradient descent. Common loss functions include mean squared error, cross-entropy, and binary cross-entropy.
Deep Learning (DL)~\cite{alom19} is a subfield of NNs that utilizes multiple hidden layers to extract hierarchical features that are useful for tasks such as pattern recognition and classification. 
Two types of supervised learning approaches for DL are Deep Neural Networks (DNNs) and Convolutional Neural Networks (CNNs).
DNNs can be applied to various tasks, including speech recognition, natural language processing, and image classification. They consist of multiple layers of artificial neurons that adjust the strengths of their connections to learn complex patterns and relationships in data.
%
CNNs are particularly suited for image processing tasks. They use a combination of convolutional layers, pooling layers, and fully connected layers to extract features from images and classify them into different categories. 
Both CNNs and DNNs have been shown to be highly effective for a variety of tasks, and continue to be an active area of research in the field of machine learning.

Homomorphic Encryption~(HE) is a type of encryption scheme that allows computations to be performed on encrypted data without decrypting it first. HE can be categorized into
three categories: Partially Homomorphic Encryption~(PHE), Somewhat Homomorphic Encryption~(SWHE) and Fully Homomorphic Encryption~(FHE). 
In PHE schemes, the homomorphic property is satisfied by only one operation, such as addition or multiplication, an unlimited number of times. SWHE supports more than an operation but a limited number of times 
Finally, FHE schemes 
allows a set of operations (usually addition and multiplication) an unlimited amount of times. 
The first FHE scheme was introduced by Gentry in 2009~\cite{gentry_fully_2009}, and it represented a significant breakthrough in the field, laying the foundation for future research. Gentry also introduced the concept of bootstrapping in FHE, which reduces the noise generated during operations and ensures a correct decryption output. However, this bootstrapping process is computationally expensive.
There are, however, several research branches on the topic of HE which are usually defined by the mathematical problems in which the security of the proposed models rely on. For instance, the FHE schemes BGV, BFV, and TFHE are based on the hardness of the Learning With Error over Rings (Ring-LWE) problem~\cite{AIHE_Lou20}. Both schemes BGV and BFV operate on integers, while  TFHE operates in binary data. Additionally, the CKKS scheme is based on a variant of the Ring-LWE problem called the approximate number field sieve, and it is meant for floating point arithmetic~\cite{cheon_homomorphic_2017}. Finally, the SWHE scheme YASHE is founded on the Short Integer Solution problem, and it also operates on integers~\cite{cryptoeprint:2013/075}. 

Several popular libraries provide implementations of the HE schemes mentioned above. For instance, the SEAL library\footnote{https://github.com/microsoft/SEAL} provides implementations of the BGV, BFV, and CKKS schemes. Earlier versions of SEAL also included an implementation of the YASHE scheme. Another library, HElib\footnote{https://github.com/homenc/HElib}, implements the BGV and CKKS schemes. Finally, the TFHE library\footnote{https://www.tfhe.com/} offers an implementation of the TFHE scheme.

\section{Methodology}

The aim of this work is to present a comprehensive and well-structured overview of the current research literature concerning the use of HE in NN for preserving data privacy. 
For this we formulated the following research question: 
``How has HE been used to preserve data privacy in NN training and inference?''. To address this broader question, four narrower sub-questions were defined:
\begin{description}
    \item[\textbf{(RQ1):}] What types of NNs are most used in studies about the use of HE to preserve data privacy in NN?
    \item[\textbf{(RQ2):}] What are the most commonly used HE schemes for preserving data privacy in NN?
    \item[\textbf{(RQ3):}] What are the major limitations identified in the literature regarding the use of HE to preserve data privacy in NNs?
   \item[\textbf{(RQ4}):] What are the most common approaches used to address the limitations of HE in NN, and what trade-offs need to be considered?
\end{description}
To achieve a transparent, replicable and complete answer to these questions, the guidelines of PRISMA methodology~\cite{prisma} were followed. The search terms ``Neural Network(s)'', ``Homomorphic'' and ``Privacy'' were defined and used in several reputable bibliographic databases. Since HE is commonly used in conjunction with other privacy-preserving techniques, we also defined a set of exclusion terms that must be omitted from the results:  ``Federated'', ``Mining'', `` Blockchain'', ``Mobile'', ``Multi-Party'', ``Edge'', ``Multiparty'', ``Distribut*'', ``Communication'', where '*' denotes any sequence of characters.

%
%
%

\begin{figure}
    \centering
    \includegraphics[width=1\textwidth]{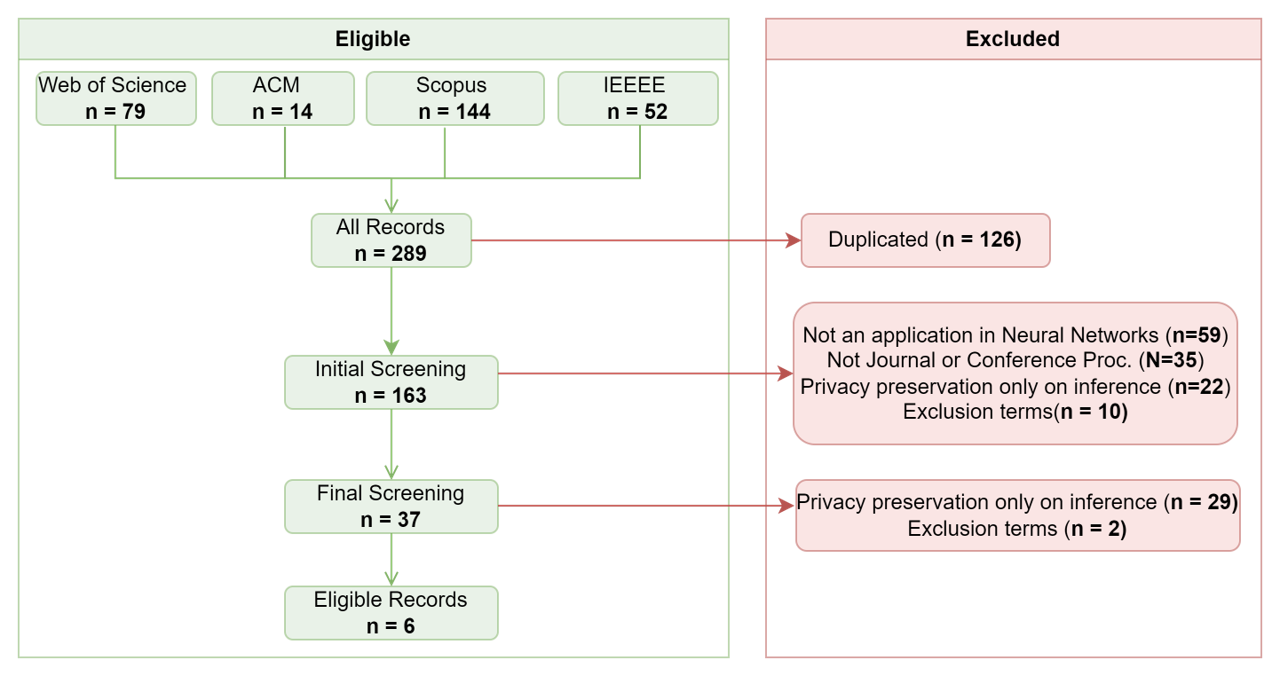}
    \caption{Eligibility process}
    \label{fig:screening}
\end{figure}

The selection criteria used in this study were clear and concise. To be included, publications had to be written in English, published in peer-reviewed journals or conference proceedings, and must discuss or evaluate the use of HE in NN training and inference. 
%
%
%
The search was then executed in four different bibliographic databases: Web of Science\footnote{https://webofscience.com/}, ACM\footnote{https://dl.acm.org/}, Scopus\footnote{https://www.scopus.com/} and IEEE Explore\footnote{https://ieeexplore.ieee.org}, and a total of 289 records were retrieved which resulted in 163 unique records. Following an initial screening of the titles and abstracts of the identified publications, a total of 126 papers were excluded. The full texts of the remaining 37 papers were then evaluated for eligibility, and only 6  were considered suitable for this study. The process described is illustrated in Figure \ref{fig:screening}.

\section{Results and Discussion}



The first known work that provides a solution for training NN models using HE was published in 2017 by Hesamifard et al.~\cite{AIHE_Hesamifard17}. 
The authors implemented NN over homomorphically encrypted data where the Sigmoid and ReLU activation functions are replaced by low-degree polynomial approximations to enable the use of NN with HE. The authors also proposed that when the ciphertext noise reaches a given threshold, it should be sent back to the client so that a fresh ciphertext can be returned, which they claim to be a more efficient alternative to bootstrapping.
Experiments were conducted on three different datasets using five NN configurations, and the reported accuracies are above 99\%, outperforming two state-of-the-art approaches. One of those approaches also used HE, but only for classification, assuming that the model was trained in plaintext. The other approach used secure multi-party computation (SMC) techniques to ensure privacy preservation in NN training and classification.
The number of communications needed for training is also referred as an advantage over the state-of-the-art SMC approach, due to the significant reduction in the required number of interactions between the server and the client.
The authors used the BGV scheme and its implementation in the HElib library.

In the following year, Ghimes et al.~\cite{AIHE_Ghimes18} implemented a NN where all data was encrypted, and all operations were performed over encrypted data using the YASHE scheme provided by the SEAL library. They utilized the simplest activation function possible, the identity function, to reduce the cost of computing over encrypted data. Their primary goal was to enable institutions to outsource NN predictions to third parties while ensuring data security and privacy. However, as recognized by the authors, their approach was not very promising, as it produced high error values and required long computation times. 
It is worth noting that the authors did not reference the previous work of Hesamifard et al. in their paper.

In 2019,  Nandakumar et al.~\cite{AIHE_Nandakumar19} 
published the first attempt to train a DNN on data encrypted using FHE, which can also be used for inference. 
Similar to Hesamifard et al.'s work, they used the HElib library to encrypt the data using the BGV scheme and to perform the computations on encrypted data. 
In this work, the authors identified two challenging steps in using HE, namely computing the activation function and its derivative, and computing the loss function and its derivative. To address this, they pre-computed these functions in a table and performed a homomorphic table lookup, an approach similar to the one proposed by Crawford et al.~\cite{10.1145/3267973.3267974}, instead of approximating them by a low-degree polynomial as Hesamifard et al. did. Their approach is faster and shallower when applicable, but can only provide a low-precision approximation of these functions. The authors used the sigmoid activation function and quadratic loss function, which have simpler derivatives, to avoid excessive table lookups.
To minimize the number of bootstrapping operations and enable parallelization of computations, a data-packing strategy was employed, leading to a significant reduction of the overall computational complexity. 
The authors conducted experiments on the MNIST dataset, and the less complex four-layer NN reported an accuracy of 96\% with an execution time of 40 minutes while using optimizations and multiple CPU threads. Although promising, the authors recognized that training data in the encrypted domain with their approach is four to five times slower than training in the plaintext domain, which shows that further research is required to improve efficiency.

After the work of Nandakumar et al., Lou et al.~\cite{AIHE_Lou20} published, in 2020, a faster alternative, that they called ``Glyph'', to train DNNs on encrypted data by combining TFHE and BGV schemes. Their approach employs a scheme switching technique that introduces only small computing overhead, but allows leveraging the use of TFHE for nonlinear activations, such as ReLU and softmax, and BGV for multiplicity-accumulations. This is possible because their scheme switching technique maps the plaintext spaces of BGV and TFHE to a common algebraic structure using natural algebraic homomorphisms, enabling, in this way, homomorphic
switching between different plaintext spaces.
The authors performed several experiments and have compared their approach with the previous one. To achieve similar accuracies, 97.8\% for the approach of Nandakumar et al. and 98.6\% for Glyph, they estimated a training time of 13.4 years for the first approach and 8 days for the second one. Overall, their experimental results show that Glyph obtains state-of-the-art accuracy, and reduces training latency by 69\% to 99\% over prior FHE-based privacy-preserving techniques on encrypted datasets.

In 2021, Onoufriou et al.~\cite{AIHE_Onoufriou21} published a study which focus on the use of HE to preserve privacy in NN training and classification for milk yield forecasting in the agri-food sector. The authors employed a one-dimensional CNN (1D CNN) with encrypted data using the CKKS scheme. Since certain operations cannot be performed on fully homomorphically encrypted data, the authors implemented a set of measures to address this issue. First, they performed the forward pass in the encrypted domain by replacing the Sigmoid activation function by a polynomial approximation proposed by Chen et al.~\cite{cryptoeprint:2018/462}, in a similar approach to the one proposed by Hesamifard et al. This helped to avoid the necessity of computing divisions in the standard Sigmoid. Then, the backward pass and weight updating were done in plaintext to reduce computational cost. 
With this approach, the authors achieved an average accuracy of 87.6\% and a Mean Absolute Percentage Error (MAPE) of 12.4\% for milk yield prediction, demonstrating that FHE allows for absolutely private predictions. However, their approach has a flaw which is acknowledged by the authors, since it does not use encrypted data in all stages of training. This limitation requires further research to be addressed. 
It is worth mentioning that for their experiments, the authors used the MS-SEAL library, and a dataset with 30 years of breeding, feeding, and milk yield data. The study showed that although FHE incurs a high spatial complexity cost, the run time remains within reasonable bounds.

In the same year, Yoo et al.~\cite{AIHE_Yoo21} proposed a novel framework, named ``t-BMPNet'', for designing a foundational model of deep learning, namely a Multilayer Perceptron Neural Network~(MLPNN), over an FHE scheme. This framework allows for training in the encrypted domain without the need to replace activation functions with polynomial approximations, unlike the approach taken by Onoufriou et al. The authors achieved this by using the TFHE scheme, which enables building complex operations over encrypted data using primitive gates and fundamental bitwise homomorphic operations. They utilized the sigmoid function as the activation function and explained how to define the four main operations needed in the encrypted domain, namely: addition, two’s complement, exponential function, and division. 
The authors conducted several experiments to validate their approach and confirmed that ``t-BMPNet'' achieved a highly accurate design of the nonlinear sigmoid function compared to other works that use polynomial approximations. However, the authors also recognized that their approach had low time performance and needs further research.

 \subsection{Discussion}

After analysing the state-of-the-art studies on the use of HE to preserve data privacy in NN training and inference, in this section we will 
discuss the main findings, considering the research questions that were defined.

\vspace{0.3cm}
\noindent \textbf{RQ1 - What types of NNs are most used in studies about the use of HE to preserve data privacy in NN? }
\vspace{0.1cm}

 Based on the analysis of the reviewed literature, it is evident that DNNs and CNNs are the most commonly used types of NNs in studies that employ HE for preserving data privacy during training and inference. Out of the six works analysed, two of them utilized CNNs, while the other five employed DNNs. In fact, in the research conducted by Lou et al.~\cite{AIHE_Lou20}, their approach was applied to both a DNN and a CNN. 

\vspace{0.3cm}
\noindent\textbf{RQ2 - What are the most commonly used HE schemes for preserving data privacy in NN?}
\vspace{0.1cm}

From our analysis of the literature, it is clear that FHE is the preferred type of scheme. 
This is not surprising, given that FHE allows for multiple operations an unlimited number of times. BGV stands out as the most commonly used in this context, with three out of the six reviewed works using it~\cite{AIHE_Hesamifard17,AIHE_Nandakumar19,AIHE_Lou20}. It is worth noting that this scheme was prevalent in the first works published, probably because an implementation of it was available at the time in the HElib library. 

More recently, other FHE schemes have been explored.  Two out of the six reviewed works employed the TFHE scheme~\cite{AIHE_Lou20,AIHE_Yoo21}, and the authors utilized the TFHE library for their experiments. By the other hand, one work uses the CKKS scheme~\cite{AIHE_Onoufriou21}.
Lou et al.~\cite{AIHE_Lou20} approach is also interesting since both BGV and TFHE are combined to leverage the best characteristics of each. Specifically, they used TFHE for nonlinear activations and BGV for multiplicity-accumulations.
Regarding SMHE, only a single work was found ~\cite{AIHE_Ghimes18}, namely the YASHE scheme (with the SEAL library being used for the experiments).

\vspace{0.3cm}
\noindent\textbf{RQ3 - What are the major limitations identified in the literature regarding the use of HE to preserve data privacy in NNs?}
\vspace{0.1cm}

One of the key limitations 
is related with the fact that HE only supports basic arithmetic operations, such as addition and multiplication, as explained in Section~\ref{sec:back}. 
Thus, other commonly used operations in NNs, such as division, exponential, comparison, and maximum, cannot be directly supported by HE or they are computationally expensive. Consequently, some restrictions are introduced on the characteristics of the NNs that can be trained with homomorphically encrypted data. For instance, activation functions like the sigmoid, which involve division, cannot be directly used. Similarly, loss functions may also be affected by unsupported operations.
Additionally, pooling techniques, such as max pooling and average pooling, which are commonly used in CNNs, can also be affected because they involve comparing and selecting the maximum or average value within a set of values, which is not possible in the majority of HE schemes.

Another significant limitation of homomorphically encrypted NNs is their computational overhead. The operations performed on encrypted data are computationally expensive, and  their complexity increase with the size of the data and the depth of the NNs. Consequently, training and inference times can become significantly longer when compared with non encrypted NN.

Finally, the use of HE with NNs can also introduce limitations on the way data is represented. HE schemes typically operate on encrypted data in a fixed-point representation, which is different from the floating-point representation used in traditional NNs. This difference can lead to numerical instability and can limit the range of values that can be represented, which may negatively affect the accuracy and performance of the model. 

\vspace{0.4cm}
\noindent\textbf{RQ4 - What are the most common approaches used to address the limitations of HE in NN, and what trade-offs need to be considered?} 
\vspace{0.1cm}

One of the most important limitations is related with the NN operations, as mentioned in the previous research question. To address this issue, most of the studied works replaced the activation and loss functions by a polynomial approximation~\cite{AIHE_Hesamifard17,AIHE_Ghimes18,AIHE_Onoufriou21}.  However, this approach has shown to negatively affect the performance of the NNs, as low-degree approximations become progressively worse for input values farther from zero. To mitigate this problem, some authors have suggested normalizing input values, but this can be difficult to control, especially for NN with several layers. To avoid using polynomial approximations, other approaches have been proposed, such as pre-computing activation and loss functions and performing homomorphic table lookups~\cite{AIHE_Nandakumar19} or implementing complex operations using fundamental bitwise operations~\cite{AIHE_Yoo21}. However, each of these approaches also has trade-offs. For instance, while the table lookup approach is faster than using polynomial approximations, it can only provide low-precision approximations for the activation and loss functions. On the other hand, implementing complex operations using fundamental bitwise operations can lead to better approximations, but it can also result in low time performance. Finally, Lou et al.~\cite{AIHE_Lou20} proposed a combined approach using the schemes TFHE and BGV, that slightly increased the accuracy of this system. 

Regarding the computation overhead limitation, 
one common strategy is to use a low degree polynomial approximation for the activation and loss functions, as explained before. This reduces the number of operations required, thereby reducing the computational overhead. Additionally, another way to reduce the computational overhead is to use NNs with fewer layers. 
In fact, most of the works reviewed have used DNN with just three layers~\cite{AIHE_Ghimes18,AIHE_Nandakumar19,AIHE_Yoo21}. 
However, some other studies have explored different NN configurations, such as varying the number of hidden layers from 1 to 5~\cite{AIHE_Hesamifard17}, or utilizing a 1D CNN~\cite{AIHE_Onoufriou21}. Since each homomorphic operation performed in encrypted data is slower than its counterpart, by reducing the layers, the number of homomorphic operations that need to be used can be reduced, which can lead to a significant reduction in the computational overhead. However, this reduction in the number of layers may also result in a reduction in the accuracy of the model. Therefore, there is a clear trade-off between model accuracy and computational overhead when using HE in NN training and classification.

In relation to the limitation which may be introduced by HE regarding the way data is represented, most of the approaches use encoding strategies to make it possible to work with low-precision integers and fixed-point representation. By the other hand, CKKS, used in the work of Onoufriou et al.~\cite{AIHE_Onoufriou21}, 
uses a floating-point representation for plaintext and ciphertext data, requiring minimal adaptations. Even though these strategies are needed to use HE in NNs, they also come with a trade-off, such as potential accuracy loss with fixed-point representation and the need for additional computation with low-precision integer representation.

\section{Conclusion}

In this work, we have reviewed and analysed the state-of-the-art on the use of HE to preserve data privacy in NN training and inference. 
Our study showed that BGV scheme is the most commonly used in this context, which highlights the trend towards the use of FHE schemes.  
The Glyph and t-BMPNet approaches stand out for leveraging the characteristics of the TFHE scheme to reduce training times while maintaining high accuracy levels, without replacing nonlinear functions with polynomial approximations.
Nevertheless, current approaches are still slow and may not provide high levels of accuracy. We believe that the selection of the most appropriate strategy depends on the specific application scenario, which may have varying requirements in terms of accuracy and computational complexity. Therefore, future research should focus on exploring the application of TFHE in more complex NNs and investigating other HE schemes. Additionally, improving computational performance through parallelization 
may be worth exploring to make HE more reliable and scalable for this type of architecture.

In summary, the use of HE for NN training is a promising approach for preserving privacy, but optimizing its accuracy and efficiency requires further research and experimentation with novel techniques and strategies.

\bibliographystyle{splncs04}
\bibliography{bibliography}

\end{document}